\documentclass[12pt]{article}
\usepackage{psfig,epsf}
\usepackage{a4wide,graphicx}
\usepackage{cite}

\begin{document}

\title{On the spin, parity and nature of the $\Xi(1620)$ resonance}

\author{
A. Ramos \\
{\small Departament d'Estructura i Constituents de la Mat\`eria,
Universitat de Barcelona,} \\
{\small Diagonal 647, 08028 Barcelona, Spain}\\
~\\
E. Oset \\
{\small Departamento de F\'{\i}sica Te\'orica and IFIC,
Centro Mixto Universidad de Valencia-CSIC,} \\
{\small Institutos de
Investigaci\'on de Paterna, Aptd. 22085, 46071 Valencia, Spain}\\
~\\
C. Bennhold \\
{\small Center for Nuclear Studies and Department of Physics,} \\
{\small The George Washington University, Washington DC 20052,  USA}
}

\date{\today}

\maketitle
\begin{abstract}

Using a unitary extension of chiral perturbation theory
with a lowest-order s-wave SU(3) chiral Lagrangian we study low-energy 
meson-baryon scattering in the strangeness $S=-2$ sector.
A scattering-matrix pole is found around 1606 MeV which corresponds to an
s-wave $\Xi$ resonance with $J^P=1/2^-$.
We identify this resonance with the $\Xi(1620)$ state, 
quoted in the Particle Data Book with $I=1/2$ but with
unknown spin and parity. The addition of the $S=-2$ state
to the recently computed $\Lambda(1670)$, $\Sigma(1620)$ and 
$N(1535)$ states completes the octet of $J^P=1/2^-$
resonances dynamically generated in this chiral unitary approach. 

\end{abstract}

Since the discovery of the first baryon resonance, the
$\Delta(1232)$, over fifty years ago, there has been a 
persistent question regarding the nature of such resonances:
Are these genuine states that appear as bare fields in the 
driving term of the meson-baryon scattering matrix?
Or can they be generated dynamically, 
i.e., by iterating an appropriate nonpolar driving term
to all orders, just like the deuteron appears as a bound
state when the nucleon-nucleon interaction is solved to all orders?
Early attempts by Chew and Low to generate the $\Delta(1232)$ by iterating
a crossed nucleon pole term were eventually overtaken by the
success of the SU(3) quark models which established the
$\Delta(1232)$ as part of the SU(3) ground-state decuplet.
The subsequent discovery of over 100 additional baryon resonances
and their mostly successful incorporation into quark models 
appeared to have settled the question in favor of treating them
as genuine fields.  However, one persistent exception for many years was
the lowest-lying $S=-1$ resonance, the SU(3) singlet $\Lambda(1405)$, which
appeared quite naturally as a dynamical pole in the $K^- p$ 
scattering matrix using a variety of approaches (see the discussion
in the Particle Data Book\cite{pdg} about the history of the $\Lambda(1405)$).
The advent of chiral Lagrangians combined with unitarization
techniques placed these efforts
on more solid theoretical grounds\cite{Kai95,Kai97,angels,joseulf}.  
Next came the suggestion that the $N^*(1535)$, which in some works 
is considered 
the $J^P=1/2^-$ chiral partner of the nucleon and hence
 would be degenerate with
the nucleon were it not for chiral symmetry breaking \cite{detar,jido2},
 can also be
generated dynamically, extending the same techniques into the $S=0$
sector\cite{siegel,Kai95,Kai97,hosaka,nieves,inoue}. 
Invoking the same framework
but extending it to higher energy, Ref.\cite{cornelius} returned
the search to the $S=-1$ sector and identified the $\Lambda(1670)$
and $\Sigma(1620)$ resonances as two additional members of the
same $J^P=1/2^-$ octet that appear naturally within the chiral unitary
framework. 

The purpose of this paper is to extend this chiral approach with unitarization
yet again to the $S=-2$ sector and demonstrate that
the last remaining member of the lowest-lying $J^P=1/2^-$ octet,
an s-wave $\Xi^*$ resonance, can indeed be found as a pole in the
appropriate meson-baryon scattering matrix.  
The experimental situation of the  $S=-2$ low-lying  $\Xi$
resonances is rather unclear. There is a well-known p-wave
state, the $P_{13}$ $\Xi(1530)$ resonance, with
$I(J^P)=1/2(3/2^+)$, which has a 4-star rating in
the Particle Data Book \cite{pdg}.  Next come the $\Xi(1620)$ and
$\Xi(1690)$ resonances which are rated with one and three
stars, respectively. While an isospin value of $I=1/2$ is quoted
for both resonances, the spin and parity of
these states have not been measured.  In the present paper we present arguments
that the $\Xi(1620)$ corresponds to the lowest-lying $J^P=1/2^-$ octet of baryon
resonances generated dynamically through multiple scattering of
meson-baryon pairs in a coupled-channels approach. 

  The states are generated in the following way. One constructs the set of coupled
channels from the octets of ground-state baryons and pseudoscalar
mesons and, using the SU(3) version of the chiral Lagrangians for
mesons and baryons \cite{Pi95,Eck95,Be95,Mei93}, one implements
any of the unitary extensions of chiral perturbation theory to
generate the scattering amplitudes connecting the various
channels. Recent unitary extensions of chiral perturbation
theory ($U\chi PT$) rely upon the inverse amplitude method of
\cite{dobado} extended to coupled channels \cite{OOP97} or the N/D
method including explicit fields of genuine resonances (which
would survive in the large $N_C$ limit) \cite{nsd}, also used in
\cite{ulfjose} to study $\pi N$ scattering.  This latter method
was also used in \cite{joseulf} to study the low-energy
interaction of antikaons with baryons and it was found to be
equivalent to the formalism used in \cite{angels} to study the
same problem using the Bethe-Salpeter equation with coupled
channels, with the difference that the loops were regularized in
\cite{joseulf} with dimensional regularization, while a cutoff
was applied in \cite{angels}.

In order to search for a $\Xi$ resonance we follow closely the
work of Ref. \cite{cornelius}, where the $\Lambda(1670)$ and
$\Sigma(1620)$ resonances were discovered in the $S=-1$ channel.
Here we focus on the $S=-2$ sector for which, as an example, the
zero-charge states of the coupled-channels framework are 
$\pi^+ \Xi^-$, $\pi^0 \Xi^0$, $\bar{K^0} \Lambda$,
$K^-\Sigma^+$, $\bar{K^0} \Sigma^0$ and $\eta \Xi^0$. 
We solve the coupled-channels Bethe-Salpeter equation for the scattering
amplitude
\begin{equation}
T = [1 - V \, G]^{-1}\, V \ ,
\label{eq:bs1}
\end{equation}
where the driving (kernel) $V$ matrix,
\begin{eqnarray} 
 V_{i j} = - C_{i
j} \frac{1}{4 f^2}(2\sqrt{s} -
M_{i}-M_{j})
\left(\frac{M_{i}+E_i}{2M_{i}}\right)^{1/2}
\left(\frac{M_{j}+E_j}{2M_{j}} \right)^{1/2}\, , \label{eq:ampl2}
\end{eqnarray}    
is obtained from the 
the chiral Lagrangian for the meson-baryon
interaction at lowest order and 
$C_{ij}$ are SU(3) coefficients.
The (diagonal) matrix $G$ in Eq.~(\ref{eq:bs1}) accounts for the loop
integral of a meson and a baryon propagator and 
depends on the regularization scale, $\mu$, and a subtraction constant for
each channel, $a_l$, that comes
from a subtracted dispersion relation. The
explicit expression of $G$ can be found in
Refs.~\cite{joseulf,cornelius}.
The regularization scale $\mu$ is of course arbitrary
but the subtraction constants depend on it.

In the present work we use the isospin basis, which 
contains the states $\pi \Xi$,  $\bar{K} \Lambda$,
$\bar{K}\Sigma$ and $\eta
\Xi$ for isospin $I=1/2$ and the states $\pi \Xi$ and $\bar{K}
\Sigma$ for isospin $I=3/2$. 
For the particular isospin $I=1/2$ case needed in the present study, the
coefficients $C_{ij}$ have the values shown in
Table~\ref{tab:table1}. We have four different subtraction
constants, namely $a_{\pi \Xi}$, $a_{\bar{K}\Sigma}$, $a_{\bar{K}
\Lambda}$ and $a_{\eta \Xi}$.
Assuming that the regularization scale $\mu$ gives the size of the
maximum momentum in
the cutoff regularization method, then the values of $a_l$ can be
deduced approximately from \cite{joseulf} and they are found to be
of the order of $-2$ when a cutoff of 630 MeV is chosen as done
in \cite{angels}.  This value for the subtraction constant is what
is called a magnitude of natural size in \cite{joseulf}, since it
corresponds to using cutoffs of the order of 1 GeV which are
considered natural in the chiral approaches. 
We point out that the choice of the coefficients $a_l$ accounts
for contributions of higher-order Lagrangians  to
the process, as shown in \cite{OOP97}.

\begin{table}[ht]
\centering \caption{\small Coefficients $C_{ij}$ of the meson
baryon amplitudes for isospin $I=1/2$ ($C_{ji}=C_{ij}$)} \vspace{0.5cm}
\begin{tabular}{l|rrrr}
        & $\pi \Xi$ &  ${\bar K} \Lambda$ &
${\bar K} \Sigma$ & $\eta\Xi$ \\
        \hline
 & & & & \\
$\pi\Xi$ & $2$ & $-{\frac{3}{2}}$ &
$-{\frac{1}{2}}$ & 0  \\
 & & & & \\
${\bar K}\Lambda$ &   &
0 & 0 &
$-{\frac{3}{2}}$  \\
 & & & & \\
${\bar K}\Sigma$ &  &  &
2 &
${\frac{3}{2}}$ \\
 & & & & \\
$\eta\Xi$ &  &  &  &
0
\end{tabular}
\label{tab:table1}
\end{table}

In the study of $S=-1$ resonances performed in \cite{cornelius}
the $a_l$ parameters were extracted by matching the results to
those of \cite{angels} and the range of values obtained, from
$-1.84$ to $-2.67$, serves as an indication for what we might assume
as reasonable natural size parameters
in the present $S=-2$ study.   
We search for poles in the second
Riemann sheet of the scattering amplitude, focussing on the elastic $\pi
\Xi \to \pi \Xi$ amplitude in the $I=1/2$ channel. 
As a trial run we set the four values of the subtraction constants
to a value of $-2$ and
we discover a pole at $1607 + i 140$
MeV. This would lead to a width around 280 MeV, unacceptably large
compared to those of the two I=1/2 resonances of interest, the
$\Xi( 1620 )$ and the $\Xi(1690)$, which are reported to be of the
order of 50 MeV or less. The mass of the particle, around 1607
MeV, would be closer to the $\Xi(1620)$ resonance.

Allowing the subtraction constants $a_l$ to change within a reasonable
natural range, we obtain the results shown in Table \ref{tab:table2}. Only
$a_{\pi \Xi}$ and $a_{\bar{K}\Lambda}$ are
varied, since we find the couplings of the resonance to the $\bar{K}\Sigma$
and $\eta \Xi$ states to be very weak and therefore the results are
insensitive to the subtraction constants corresponding to these two
channels.
The values of the couplings, calculated from the residue of 
the diagonal scattering amplitudes \cite{cornelius},
are also shown in Table~\ref{tab:table2}. 

\begin{table}[ht]
\centering \caption{\small Resonance properties for various sets
of subtraction constants } \vspace{0.5cm}
\begin{tabular}{l|c|c|c|c|c|}
 & Set 1 & Set 2 & Set 3 & Set 4 & Set 5 \\
\hline
$a_{\pi\Xi}$          & $-2.0$ & $-2.2$ & $-2.0$ & $-2.5$ & $-3.1$ \\
$a_{{\bar K}\Lambda}$ & $-2.0$ & $-2.0$ & $-2.2$ & $-1.6$ & $-1.0$ \\
$a_{{\bar K}\Sigma}$  & $-2.0$ & $-2.0$ & $-2.0$ & $-2.0$ & $-2.0$ \\
$a_{\eta\Xi}$         &  $-2.0$ & $-2.0$ & $-2.0$ & $-2.0$ & $-2.0$ \\
\hline
$\mid g_{\pi\Xi}\mid^2$          & 8.7  & 7.2  & 7.4  & 7.2  & 5.9 \\
$\mid g_{{\bar K}\Lambda}\mid^2$ & 5.5  & 4.6  & 4.2  & 5.8  & 7.0 \\
$\mid g_{{\bar K}\Sigma}\mid^2$  & 0.68 & 0.59 & 0.54 & 0.74 & 0.93 \\
$\mid g_{\eta\Xi}\mid^2$         & 0.36 & 0.27 & 0.38 & 0.14 & 0.23 \\
 \hline
$M$ & 1607 & 1597 & 1596 & 1604 & 1605 \\  
$\Gamma/2$ & 140 & 117 & 134 & 98 & 66 
\end{tabular}
\label{tab:table2}
\end{table}

The second and third columns in Table~\ref{tab:table2} show that 
a change of 10\% in the subtraction constants
$a_{\pi \Xi}$ and $a_{\bar{K}\Lambda}$ modifies the mass of the
resonance only slightly but has a larger influence on the width.
Investigating the dependence of the results on the values of these
two subtraction constants we observe that the mass of the resonance is
confined to a range around 1600 MeV. The width, on the other hand,
can be reduced considerably
by a simultaneous increase of the strength of  $a_{\pi
\Xi}$ and a decrease of $a_{\bar{K}\Lambda}$, while
keeping both of them negative and still reasonably close
to the reference value of $-2$. In the last 
column we see that the width can be reduced to 130
MeV with acceptable values for the coefficients. While this width
might still appear as grossly overestimating the experimental ones,
we show below that this is not the case.

\begin{figure}[tbp]
\begin{center}
\includegraphics[width=7cm,angle=-90]{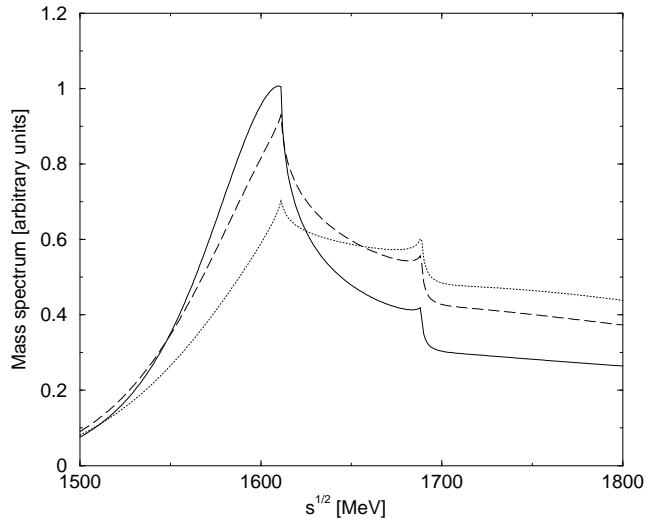}
\caption{The $\pi\Xi$ invariant mass distribution as a function of the
center-of-mass energy, for several sets of subtraction constants.
Solid line: $a_{\pi \Xi}=-3.1$ and $a_{\bar{K}\Lambda}=-1.0$;
Dashed line: $a_{\pi \Xi}=-2.5$ and $a_{\bar{K}\Lambda}=-1.6$;
Dotted line: $a_{\pi \Xi}=-2.0$ and $a_{\bar{K}\Lambda}=-2.0$. The
value of the remaining two other subtraction constants, $a_{{\bar K}\Sigma}$
and $a_{\eta\Xi}$, is fixed to $-2.0$ in all curves.}
\label{fig:masspixi}
\end{center}
\end{figure}

Since the 
$\Xi(1620)$ resonance decays only into 
 $\pi \Xi$ final states, it is experimentally visible through
the $\pi \Xi$ invariant mass distribution 
in reactions leading, among others, to $\pi$ and $\Xi$
particles. Our calculated distribution, displayed
in Fig.~\ref{fig:masspixi}, 
shows a very interesting feature, namely a smaller
apparent width  compared to the one obtained
at the pole position. 
For instance, for the values of the
subtraction constants in the last column of Table~\ref{tab:table2} we
see in Fig.~\ref{fig:masspixi} (solid line) an
apparent Breit-Wigner width of around 50 MeV and a shape for the
distribution which resembles the experimental peaks observed.
This well-known phenomenon, usually referred to as Flatt\'e effect
\cite{flatte}, is due to the presence of a
resonance just below the threshold of a channel to which the
resonance couples very strongly. In our case the
 $\bar{K} \Lambda$ channel opens at 1611 MeV and,
as shown in Table~\ref{tab:table2}, the resonance couples
very strongly to that state.
What actually happens is that at an invariant energy close
to the resonance mass the amplitude is given essentially by the
inverse of the resonance width. As soon as the
threshold is crossed, the new channel leads to 
an additional energy-dependent contribution for the width which
grows very rapidly with increasing energy. This produces a fast
fall-off for the amplitude, leading to an apparent width much smaller
than the actual width at the pole. This phenomenon has been observed,
e.g., in the case of the $a_0(980)$ meson
resonance as discussed in Refs.~\cite{OOP97,nsd}.

The question now arises which of the two $I=1/2$ candidates
should be identified with the resonance obtained here. 
The value found for the mass of the state would suggest identification
with the $\Xi(1620)$.  The experimental situation regarding 
 this 1-star resonance is the following: Three experiments see the
resonance in the $\Xi
\pi$ spectrum in reactions where $K^- p$ goes to three or four
particles in the final state, including $\Xi \pi$
\cite{ross,briefel,debellefon}. Taking into account the experimental
errors of all experiments, the resonance mass is found in the range
$1600-1640$ MeV and the width in the range $15-50$ MeV.

The $\Xi(1690)$ state is better known and is rated as a 3-star
resonance.  Even if the spin and parity are unknown, there is far
more information available for this resonance than for the $\Xi(1620)$
\cite{pdg}. 
Ref. \cite{dionisi} gives ratios of partial decay widths having
sufficient accuracy for us to draw conclusions from
the properties of the $\Xi$ resonance found in this work.
We therefore investigate whether the parameters of the theory 
provide enough flexibility
to produce a pole with a real part closer to 1690 MeV,              
since the results of Table~\ref{tab:table2} show that, by decreasing the
size of $a_{\pi \Xi}$
or $a_{\bar{K}\Lambda}$, one increases the mass of the resonance.
However, the presence of the $\bar{K} \Lambda$ threshold leads to
mass values that stabilize around the cusp of this threshold
for a certain range of the parameters.
Continuing to change the $a_l$ parameters beyond this range 
leads not to an increase of the resonance mass but 
to a disappearance of the pole --and with it the resonance--
entirely.
The above argument clearly favors identifying 
the resonance found here with the $\Xi(1620)$ state.

The other argument in favor of the $\Xi(1620)$ assignment is the
following:  The results of Table~\ref{tab:table2} show that the
resonance couples strongly to the $\pi
\Xi$ and the $\bar{K}\Lambda$ channels but very weakly to $\bar{K}\Sigma$
and $\eta \Xi$. This is opposite to the observed properties of
the $\Xi(1690)$ resonance, for which Ref.~\cite{dionisi} gives a ratio
of branching ratios for $\bar{K}\Sigma$ to $\bar{K}\Lambda$ around
3 and for $\pi \Xi$ to $\bar{K}\Sigma$ of less than 0.09. In our  
opinion, this
argument rules out identifying the resonance found here
with the $\Xi(1690)$ state.

We also point out here that using QCD sum rules one
obtains a $J^P=1/2^-$ octet of excited baryons where the $\Sigma$
and the $\Xi$ states appear degenerate and with a mass around 1620
MeV \cite{jido}. Furthermore, the lowest $\Xi$ resonances have been
seen on the lattice, however, these results do not
yet allow any quantitative conclusions\cite{frank}.  

In summary, we find a $\Xi$ resonance with an energy around 1606
MeV and a width at the pole position around 100 MeV.
Due to a significant threshold
effect the apparent (Breit-Wigner) width is much smaller and compatible with
experimental findings quoted in
\cite{pdg} for the $\Xi$ resonances in question. 
Of the two relevant $\Xi$ states with isospin 1/2, the 
1-star rated $\Xi(1620)$ and the 3-star rated $\Xi(1690)$,
we argued that the $\Xi(1690)$ must be ruled out
because, on the one hand, it was
impossible within our approach to find a pole with an energy close
to 1690 MeV and, on the other hand, there are large
disagreements between the resonance couplings to meson-baryon
states found here and the measured partial decay widths to those states
\cite{dionisi}.   
The findings presented here indicate that overwhelming evidence 
supports the assignment of the quantum numbers 
$J^P=1/2^-$ to the $\Xi(1620)$ resonance.
Thus, the computation within the chiral unitary approach of the
$\Xi(1620)$, along
with its partners, the $N^*(1535)$, the $\Lambda(1670)$ and
the $\Sigma(1620)$, plus the $\Lambda(1405)$ which largely accounts
for the SU(3) singlet, completes the $J^P=1/2^-$
nonet of dynamically generated s-wave resonances.
Clearly, the special nature of the resonance discussed here
calls for renewed experimental
efforts, especially for the 1-star rated $\Xi(1620)$, 
 $\Sigma(1620)$ states.

Since the SU(3) octet of low-lying $J^P=1/2^-$
resonances represents the chiral partner of the 
$J^P=1/2^+$ ground-state octet, the description
of these states demonstrates the extraordinary
power of the chiral unitary approach.
That these states should appear especially
with as simple a driving term as provided by
the first-order chiral Lagrangian with very few open 
parameters is truly remarkable and shows that the nature of these states is well
represented by a cloud of meson baryon components. This would be in principle
different to other excited states in higher partial waves or higher excitation
energy which might be better repesented in terms of their quark constituents.

\subsection*{Acknowledgments}

E. O. and C. B. wish to acknowledge the hospitality of the
University of Barcelona and A. R. and C. B. that of the University
of Valencia, where part of this work was done.  This work is also
partly supported by DGICYT contract numbers BFM2000-1326,
PB98-1247, by the EU-TMR network Eurodaphne, contract no.
ERBFMRX-CT98-0169, and by the US-DOE grant DE-FG02-95ER-40907.


\begin{thebibliography}{99}
\bibitem{pdg} D. E. Groom {\it et al.}, The European Physical Journal C15,
1 (2000)
\bibitem{Kai95} N. Kaiser, P.B. Siegel and W. Weise, Nucl. Phys. A594,
325 (1995)
\bibitem{Kai97} N. Kaiser, T. Waas and W. Weise, Nucl. Phys. A612,
297 (1997)
\bibitem{angels} E. Oset and A. Ramos, Nucl. Phys. A635, 99 (1998)
\bibitem{joseulf} J.A. Oller and U.G. Meissner, Phys. Lett. B500,
263 (2001)
\bibitem{detar} C. De Tar and T. Kunihiro, Phys. Rev. D39, 2805 (1989)
\bibitem{jido2} D. Jido, M. Oka and A. Hosaka, Phys. Rev. Lett. 80, 448 
(1998)
\bibitem{siegel} N. Kaiser, P.B. Siegel and W. Weise, Phys. Lett. B362,
23 (1995)
\bibitem{hosaka} J.C. Nacher, A Parre\~no, E. Oset, A. Ramos, A. Hosaka and M.
Oka, Nucl. Phys. A678, 187 (2000)
\bibitem{nieves} J. Nieves, E.R. Arriola, Phys. Rev. D64, 116008 (2001)
\bibitem{inoue} T. Inoue, E. Oset and M.J. Vicente Vacas, Phys. Rev.
C65, 035204 (2002)
\bibitem{cornelius} E. Oset, A. Ramos and C. Bennhold, Phys. Lett. B537,
99 (2002)
\bibitem{Pi95} A. Pich, Rep. Prog. Phys. 58, 563 (1995)
\bibitem{Eck95} G. Ecker, Prog. Part. Nucl. Phys. 35, 1 (1995)
\bibitem{Be95} V. Bernard, N. Kaiser and U.G. Meissner, Int. J. Mod. Phys.
E4, 193 (1995)
\bibitem{Mei93} U.G. Meissner, Rep. Prog. Phys. 56, 903 (1993)
\bibitem{dobado}  A. Dobado and J.R. Pelaez, Phys. Rev. D56, 3057 
(1997)
\bibitem{OOP97} J.A. Oller, E. Oset and J.R. Pelaez, Phys. Rev. D59,
074001 (1999); (E) Phys. Rev. D60, 099906 (1999)
\bibitem{nsd} J.A. Oller and E. Oset, Phys. Rev. D60, 074023 (1999)
\bibitem{ulfjose} U.G. Meissner and J.A. Oller, Nucl. Phys. A673, 311 
(2000)
\bibitem{flatte} S.M. Flatt\'e, Phys. Lett. B63, 224 (1976)
\bibitem{ross} R.T. Ross et al., Phys. Lett. B38, 177 (1972)
\bibitem{briefel} E. Briefel et al., Phys. Rev. D16, 2706 (1997)
\bibitem{debellefon} A. de Bellefon et al., Nuovo Cimento A28,
289 (1975)
\bibitem{dionisi} C. Dionisi et al., Phys. Lett. B80, 145 (1978)
\bibitem{jido} D. Jido and M. Oka, hep-ph/9611322
\bibitem{frank} W. Melnitchouk et al., hep-lat/0202022

\end{thebibliography}
\end{document}